\documentclass[sigconf,natbib=true,anonymous=false,authordraft=false,nonacm=true]{acmart}
\AtBeginDocument{%
  \providecommand\BibTeX{{%
    \normalfont B\kern-0.5em{\scshape i\kern-0.25em b}\kern-0.8em\TeX}}}

\copyrightyear{2023}
\acmYear{2023}

\usepackage{amsfonts}
\usepackage{graphicx}
\usepackage{subcaption}
\usepackage{mwe}
\usepackage{booktabs}
\usepackage{multirow}
\usepackage{enumitem}
\usepackage{todonotes}
\usepackage{tabularray}

\usepackage[T1]{fontenc}
\usepackage[utf8]{inputenc}
\usepackage{CJKutf8}

\usepackage{pifont}
\newcommand{\cmark}{\ding{51}}%
\newcommand{\xmark}{\ding{55}}%

\begin{document}

\title{Synthetic Cross-language Information Retrieval Training Data}

\author{James Mayfield}
\affiliation{%
  \institution{HLTCOE, Johns Hopkins University}
  \city{Baltimore, MD}
  \country{USA}
}
\email{mayfield@jhu.edu}

\author{Eugene Yang}
\affiliation{%
  \institution{HLTCOE, Johns Hopkins University}
  \city{Baltimore, MD}
  \country{USA}
}
\email{eugene.yang@jhu.edu}

\author{Dawn Lawrie}
\affiliation{%
  \institution{HLTCOE, Johns Hopkins University}
  \city{Baltimore, MD}
  \country{USA}
}
\email{lawrie@jhu.edu}

\author{Samuel Barham}
\affiliation{%
  \institution{HLTCOE, Johns Hopkins University}
  \city{Baltimore, MD}
  \country{USA}
}
\email{samuel.barham@jhuapl.edu}

\author{Orion Weller}
\affiliation{%
  \institution{Johns Hopkins University}
  \city{Baltimore, MD}
  \country{USA}
}
\email{oweller@cs.jhu.edu}

\author{Marc Mason}
\affiliation{%
  \institution{HLTCOE, Johns Hopkins University}
  \city{Baltimore, MD}
  \country{USA}
}
\email{mmason8@jhu.edu}

\author{Suraj Nair}
\affiliation{%
  \institution{University of Maryland}
  \city{College Park, MD}
  \country{USA}
}
\email{srnair@umd.edu}

\author{Scott Miller}
\affiliation{%
  \institution{ISI, University of Southern California}  
  \city{Boston, MA}
  \country{USA}
}
\email{smiller@isi.edu}

\renewcommand{\shortauthors}{Mayfield et al.}
\newcommand{\jhpolo}[0]{JH-POLO\xspace} %
\newcommand{\hcthree}[0]{HC3\xspace} %
\newcommand{\anonregimen}[0]{JH-POLO\xspace} %
\newcommand{\msmarco}[0]{MS~MARCO\xspace}

\begin{abstract}
  A key stumbling block for neural cross-language information retrieval (CLIR) systems
  has been the paucity of training data.
  The appearance of the \msmarco monolingual training set led to significant advances in the state of the art in neural monolingual retrieval.
  By translating the \msmarco documents into other languages using machine translation,
  this resource has been made useful to the CLIR community.
  Yet such translation suffers from a number of problems.
  While \msmarco is a large resource, it is of fixed size;
  its genre and domain of discourse are fixed;
  and the translated documents are not written in the language of a native speaker of the language,
  but rather in translationese.
  To address these problems,
  we introduce the \jhpolo CLIR training set creation methodology.
  The approach begins by selecting a pair of non-English passages.
  A generative large language model is then used to produce an English query for which the first passage is relevant
  and the second passage is not relevant.
  By repeating this process,
  collections of arbitrary size can be created in the style of \msmarco
  but using naturally-occurring documents in any desired genre and domain of discourse.
  This paper describes the methodology in detail,
  shows its use in creating new CLIR training sets,
  and describes experiments using the newly created training data.
\end{abstract}

\ccsdesc[500]{Information systems~Multilingual and cross-lingual retrieval}
\ccsdesc[300]{Information systems~Document collection models}

\keywords{cross-language information retrieval, CLIR, synthetic training data. domain shift, GPT-3}

\maketitle

\section{Introduction}

As with many other human language technologies,
neural models have recently achieved state-of-the-art performance in monolingual ad hoc information retrieval (IR).
A key enabler of these advances has been the appearance of large IR training sets
such as \msmarco~\cite{msmarco}.
\msmarco was developed by mining Bing query logs to identify,
for each query, a relevant and a non-relevant document
drawn from the Bing index.
This makes \msmarco well-suited to training IR systems for web-style queries
where the documents are English webpages.
It is less well-suited to other document languages, query styles and document genres as \citet{promptagator} demonstrate.
Nonetheless, \msmarco has been the basis for much of the improvement in IR
achieved by neural systems.

In cross-language information retrieval (CLIR)
there has been no resource comparable to \msmarco.
A number of CLIR collections are available.
HC4\cite{lawrie2022hc4}\footnote{\url{https://github.com/hltcoe/hc4}} and
TREC NeuCLIR~1~\cite{neuclirOverview22trec}\footnote{\url{https://neuclir.github.io/neuclir1.html}}
are high-quality ad hoc CLIR collections,
but are too small to serve as training data for a neural system.
Collections such as CLIRMatrix\cite{sun-duh-2020-clirmatrix},\footnote{\url{https://github.com/ssun32/CLIRMatrix}}
XOR-QA\cite{asai-etal-2021-xor},\footnote{\url{https://github.com/AkariAsai/XORQA}}
and MIRACL\cite{zhang2022making}\footnote{\url{https://github.com/project-miracl/miracl}}
cover numerous languages,
but like \msmarco are focused on question answering and are biased towards Wikipedia articles.
Their relevant documents are also not paired with non-relevant counterparts.

Given the lack of appropriate training sets for ad hoc CLIR,
the research community has used machine translation
to translate the \msmarco documents into other languages.
This has resulted in collections such as mMARCO\cite{mmarco}, and
NeuMARCO\footnote{\url{https://ir-datasets.com/neumarco.html}}
of the same size as \msmarco
with queries in English
and documents in another language.
Using these resources,
neural systems have achieved state-of-the-art CLIR performance.

Yet such translated training collections suffer from a number of problems.
While \msmarco is a large resource, it is of fixed size;
thus, the amount of available training data is limited.
More importantly, the genre and domain of discourse of the collection are fixed;
documents are drawn from the Bing index
and do not include, for example, informal communications such as email and Tweets.
In addition, the translated documents are not written by a native speaker of the language,
but rather suffer from a phenomenon known as \textit{translationese}~\cite{volansky2015features}:
translation artifacts that have been shown to affect
cross-language transfer performance~\cite{artetxe2020translation}.
Furthermore, \msmarco is available only for research purposes,\footnote{\url{https://microsoft.github.io/msmarco/}}
so commercial systems and other non-research applications cannot make use of it.

To address these problems,
we introduce the \jhpolo
training set creation methodology.
\jhpolo starts with a pair of non-English passages.
These passages can be written by native speakers of the language,
and can be drawn from any genre or domain.
Thus, a collection generated using the \jhpolo methodology
can be tailored to any desired retrieval setting.

Once a passage pair has been selected,
an English query is automatically generated for which one passage of the pair is relevant
and the other passage is not.
We use English as the query language to match the available CLIR test collections.
This creates an \msmarco-style training example
comprising a query, a relevant passage, and a non-relevant passage.
A  generative large language model (LLM) such as GPT-3~\cite{gpt3} is used to produce the English query.
By repeating this process,
a training collection of arbitrary size can be created.

This paper describes the \jhpolo methodology in detail,
shows its use in creating new CLIR training sets,
and describes experiments that demonstrate the efficacy of the approach.

We make the following contributions:
\begin{itemize}
    \item We show that it is possible to generate a viable large CLIR training set
    automatically using only a target document collection and a generative LLM.
    To our knowledge, this is the first automatically generated CLIR training collection
    that uses natively-written passages.
    \item We show that negative training examples can be selected \textit{before} generating the retrieval query to which they are not relevant,
    thereby allowing some control over the difficulty of negative examples
    in the generated collection.
    \item We show that training using the \jhpolo methodology is comparable to using machine-translated \msmarco data
    when the documents to be searched are similar to the web documents used by \msmarco documents,
    and more effective than training exclusively on \msmarco
    when the domain or genre of the evaluation document collection
    deviates from that of the \msmarco documents.
\end{itemize}

\begin{figure*}[!t]
    \centering
    \includegraphics[width=\linewidth]{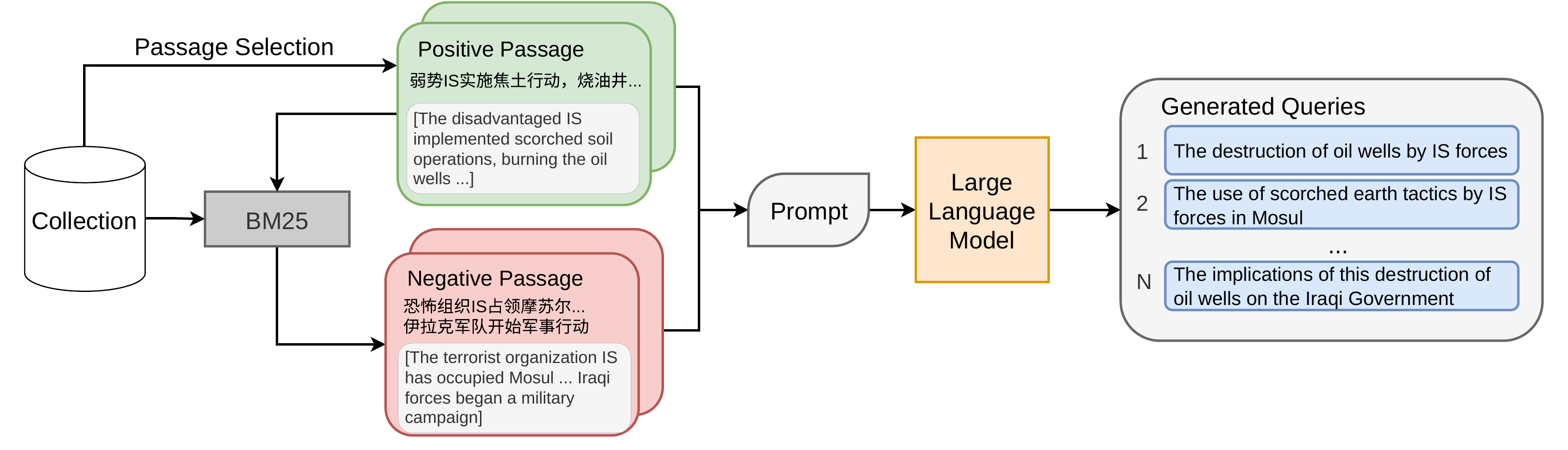}
    
    \caption{
    A depiction of the basic \jhpolo methodology. A target language passage (Chinese in this example, translated into English for convenience) is selected randomly from the target passage collection, and BM25 retrieval is used to identify a related passage. The two Passages are presented to a large language model, which is then prompted to generate queries for which one passage is relevant and the other is not.
    }
    \label{fig:method}
\end{figure*}

\section{Background}

\subsection{Cross-Language Information Retrieval}

When moving from monolingual IR to CLIR,
there is the added complexity of crossing the language barrier
between the query expression and the document expression.
One popular approach is to use a Machine Translation (MT) system
to translate either the queries or the documents,
to achieve a monolingual space where a monolingual IR system can be used~\cite{zhou2012translation, nie2010cross, galuvsvcakova2021cross}.
Another approach generates dense representations of queries and documents.
Matching queries to documents happens in a shared multilingual vector space;
this approach is popularly known as {\em dense retrieval}.
Pre-BERT~\cite{bert} dense retrieval models used non-contextualized cross-language word representations
to perform CLIR matching~\cite{yu2020study, li2018learning, gupta2017continuous}.
The adoption of large multilingual pretrained language models (mPLMs) such as mBERT~\cite{bert} and XLM-R~\cite{conneau2019xlmr}
led to dense neural CLIR systems that 
use contextualized representations for matching~\cite{shi2021cross, litschko2021cross, li2021learning, colbertx}.
Dense retrieval models for CLIR now rely heavily on mPLMs as the backbone of the models. 
However, \citet{litschko2021cross} demonstrate that performance using an off-the-shelf mPLM for CLIR is suboptimal.
While sufficient training data is
available to fine-tune an English system in the form of \msmarco~\cite{msmarco},
analogous CLIR data is not natively available.
Translated versions of \msmarco,
where the MS MARCO documents are replaced with machine translation output,
have been used to fill this gap~\cite{mmarco, colbertx}.

This paper focuses on an alternative approach to fine-tuning CLIR systems.
We explore the synthetic generation of queries from passages selected from a target document collection.
Rather than effectiveness being dependent on the quality of an mPLM or the quality of machine translation,
in this approach effectiveness is dependent on the ability of a generative LLM to produce effective training examples.

Dense Passage Retrieval (DPR)~\cite{dpr} and ColBERT~\cite{colbert}
are two of the most commonly studied and highest performing dense retrieval models. 
DPR computes the similarity of the query classification (CLS) token and the CLS token of each document.
ColBERT computes similarities between each pair of query and document tokens and scores a document by the sum of the maximum similarity (MaxSim) of each query token~\cite{colbert}. 
Compared to other neural reranking models such as a cross-encoder~\cite{nogueira2019passage}, 
dense retrieval models limit ranking latency by separating query and document transformer networks to support offline indexing.

DPR-X~\cite{mrtydi,C3,yang2022adapter} and ColBERT-X~\cite{colbertx} are the CLIR counterparts of DPR and ColBERT. 
Both use an mPLM as the underlying language model for crossing the language barrier. 
Exploiting both multilinguality and improved pre-training from XLM-R~\cite{conneau2019xlmr},
DPR-X and ColBERT-X seek to generate similar contextual embeddings for terms with similar meanings,
regardless of their language.
These are the two retrieval models featured in our experiments.

\subsection{LLMs and Retrieval}
Language models are now tightly integrated with information retrieval systems.
These combined systems are used for a broad range of knowledge-intensive problems,
including open-domain question answering~\cite{izacard2020leveraging,izacard2022few},
conversational assistants~\cite{shuster2020multi,shuster2022blenderbot}, fact-checking~\cite{thorne2018fever,thorne2018automated},
and even improving language modeling itself~\cite{lewis2020retrieval,borgeaud2022improving}. 

At times these systems are simply combinations of separate processes~\cite{izacard2020leveraging,petroni2020kilt,glass2022re2g},
while other times they are trained end-to-end from retrieval to the downstream task~\cite{lee2021you,izacard2022few,jiang2022retrieval}.
Due to the size of LLMs, they are typically used as separate components,
with retrieval results passed to the LLM~\cite{glass2022re2g,si2022prompting,kasai2022realtime}.
A nascent line of work has even proposed ignoring retrieval entirely and using LLMs to generate a relevant document in lieu of search~\cite{yu2022generate}.
In contrast to much of the research cited in this section,
our work aims to use LLMs to improve IR models,
rather than using retrieval to improve LLMs on NLP tasks.

\subsection{Synthetic Query and Document Generation}

Using LLMs to improve IR models through synthetic data generation has also been a burgeoning area of interest~\cite{wang2021gpl,schick2020few,schick2021generating,rosenbaum2022clasp,honovich2022unnatural}.
A prominent early example is the doc2query~\cite{doc2query} family of algorithms,
which supports the generation of a query that is relevant to a given document
and which is then appended to it as a form of document expansion.
As language models have grown in size and ability~\cite{gpt3},
there has been a surge of interest in this topic.
HyDE~\cite{hyde} uses LLMs to generate a synthetic document
that is then used as a query,
while the InPars algorithms~\cite{bonifacio2022inpars,jeronymo2023inpars,boytsov2023inpars}
and PROMPTAGATOR~\cite{promptagator}
use LLMs to generate queries given document,
in the reranking and end-to-end settings respectively.
These works differ in how they prompt the LLMs:
PROMPTAGATOR uses a different prompt template for each dataset
and only shows relevant few-shot examples
(i.e., what the LLM should generate)
while InPars also uses non-relevant few-shot examples
(i.e., what not to generate).

Despite the plethora of recent research in creating synthetic training data for IR,
to date, and with a few exceptions (e.g., HyDE~\cite{hyde}),
most work has focused on the English language.
This leaves it unclear how LLMs can be used to train translingual or multilingual IR systems.

\section{\jhpolo Method}

Generation of a single training example starts with the selection of two passages.\footnote{Full-length documents could exceed length limits imposed by the LLM.}
A generative LLM is given these passages
and prompted to compose an English query for which one passage is relevant
and the other is not.
This process is repeated to generate as many training examples as desired.

This method has two significant advantages:
\begin{enumerate}

    \item It ensures that the passages are naturally-occurring text
    selected from the language, genre and domain of interest.
    Use of \msmarco for CLIR has relied on machine translation of the \msmarco document collection,
    which exhibits artifacts of machine translation.
    Furthermore, there is no way to alter the characteristics of the document collection
    underlying \msmarco.

    \item It exploits a generative LLM's strength, which is generating short English\footnote{At this writing the major generative LLMs focus on English.} texts.
    LLMs can struggle when trying to generate a long document.
    Its capabilities in languages other than English are also inconsistent.
    By generating short English queries these problems with LLMs are ameliorated.
    
\end{enumerate}

Figure~\ref{fig:method} is a pictorial representation of the \jhpolo process. Section~\ref{sec:docselection} describes
the left side of the figure, while Section~\ref{sec:prompt} describes the right side of the figure.

\subsection{Passage Selection}
\label{sec:docselection}

Choosing passages at random would be a simple way to select two passages
for use in query generation.
However, doing so would almost always select two passages with no topic overlap.
Any system trained using such pairs would have a difficult time
distinguishing passages with a high topic overlap at test time.
We would like our training data set to include related passages
that exhibit significant overlap with a relevant passage
but are not themselves relevant.
We hypothesize that the closer the content of the two passages,
the more useful the pair will be for training. 
There are a number of ways to choose two related passages;
these include:
\begin{itemize}
    \item Use an existing document collection and passage pairs.
    \msmarco is the obvious target here;
    it has passages, and topics with an example of a relevant passage and a non-relevant passage for each topic.

    \item Use an existing ad hoc IR collection.
    For example the TREC NeuCLIR track\footnote{\url{https://neuclir.github.io/}} provides English topics with documents in Chinese, Persian, or Russian.
    One way to select a pair is to use the relevance judgments (qrels)
    to select two passages, one from a randomly chosen judged relevant document
    and the other from a randomly chosen judged non-relevant document for a given topic.
    This does not guarantee that the same relevance judgments will apply to the selected passages,
    but it is likely that those passages will be related but not identical.
    Alternatively, one could perform retrieval on the original queries,
    and use the ranked results to select two top-scoring passages.

    \item Use a collection with relatedness links.
    One could for example select two linked Wikipedia articles,
    or two versions of a single Wikipedia article from two different dates.

    \item
    Select a passage at random, or one returned by a query,
    and use that entire passage as a query.
    Select the top retrieved passage whose BM25 retrieval score
    is at least a fixed threshold away from the score of the query passage
    as the negative passage.
\end{itemize}
The last approach is the one explored in this paper.
By requiring at least some separation in the BM25 scores of the two passages,
we ensure that the two passages contain some different information.
We also require that the passages do not come from the same underlying source document.
Different genres may also necessitate additional requirements
to ensure the selection of useful training pairs.
For instance, we examine the longest common substring between two passages sourced from informal communications;
the selected passage must contain both twenty characters and 40\% of its total characters outside of that common substring.

\subsection{Prompt Specification}
\label{sec:prompt}

Unlike pre-trained language models that are routinely fine-tuned
to adapt them to new genres, domains, or tasks,
the common and economic way to use a generative LLM such as GPT-3
is to engineer a prompt to guide the desired generation.
We experimented with a variety of prompts
with the goal of creating suitable CLIR queries.
Such a prompt must:
\begin{itemize}
    \item contain the text of each of the passages.
    \item indicate what type of output is required.
    We would like to produce multiple output queries for each prompt
    to reduce the overall cost of building the collection.
    \item ensure that the generated queries are written in English
    regardless of the language of the passages.
    \item communicate what is meant by relevance.
    \item require that one of the passages is relevant to the output query and the other is not.
\end{itemize}

\begin{figure*}[tb]
\begin{verbatim}This is document A: <<{first}>>
This is document B: <<{second}>>

I am an analyst writing a report. Only one of the documents will help me write my report.  For each
document, describe in English, one per line, five things my report might be about for which that
document will help me write my report and the other document will not help me write my report.\end{verbatim}
  \caption{GPT-3 prompt used to create the training examples reported in this paper}
  \label{fig:prompt}
\end{figure*}

Figure~\ref{fig:prompt} shows the basic prompt we used to create the training collections described in this paper.
Here, \{first\} and \{second\} are replaced with the complete text of the first and second passages.
The prompt requests five outputs for each passage,
requires that the output is in English,
and stipulates that one passage must be relevant and the other not relevant.
Relevance is defined relative to an analyst writing a report;
a passage is relevant if it helps the analyst write the report,
and not relevant if it does not.
The topic of the report is not specified in the prompt;
the LLM is free to invent any report topic it likes.
Thus the output query can be on any suitable topic.

We experimented with few-shot prompts that included sample outputs.
These prompts had two problems. 
First, they increased the length of the prompt,
increasing the cost of the request, which for GPT-3 is dependent on the sum of the lengths of the input and output.
Second, there was occasionally bleed-through of the topics of the sample outputs
into the queries produced.
As a result, we restricted our attention to zero-shot prompts
that relied purely on description of the desired output.

\subsection{Crossing the Language Barrier}
\label{sec:barrier}

We use GPT-3 Davinci-3\footnote{\url{https://beta.openai.com/docs/models/gpt-3}} as our large language model for two reasons.
First, its input buffer is 4000 tokens,
allowing us to include passages of up to about 550 words of Chinese,
260 words of Russian,
or 370 words of Persian,
while allowing an additional 100 or so tokens of English for the prompt.
Second, Davinci is far more capable with languages other than English
than are the lesser GPT-3 models.
We present non-English passages to GPT-3 with no indication of what language they are written in;
the prompt indicates only that they are `documents.'
Davinci seems to handle other languages with ease;
the lesser models do not.

The ability of GPT-3 Davinci-3 to handle languages other than English
varies dramatically by language~\cite{gpt3}.
If GPT-3 is unable to handle a given foreign language well,
an alternative is to use machine translation to produce English versions of the documents.
Then an English-only process is applied to these translations.
This approach relies on document relevance not changing much when a document is translated.
This is plausible,
although the claim remains to be proven. 
It should be noted that while the LLM would process the translated documents in this case,
the CLIR fine-tuning would 
continue to use the original natively written documents.

\subsection{Failure Modes}
We have identified four categories of error most commonly seen in \jhpolo output. The following describes them in detail.

\textbf{Underspecification.} This occurs when the query could refer to something in the passage,
    but could just as easily refer to many other things completely unrelated to the passage.
    For example, ``The emergence of images in the media related to the leak''
    could refer to any of a number of instances of leaked documents.
    This failure mode can be thought of as inadequate inclusion of context in the query.
    Despite the apparent problem,
    we believe underspecified queries are less damaging as training data than other categories of errors because the negative passage is still not
    relevant to an underspecified query.

\textbf{Overspecification.} This occurs when the selected passage is relevant to the query,
    but no other passage is likely to be relevant.
    For example, ``The arrest of Moise’s bodyguards and 3 security personnel''
    describes one very specific facet of an event,
    one that will likely be found in very few of the passages about the event.
    This failure mode frequently occurs when numbers are part of the generated query,
    because it limits the query to a particular instance of a topic.
    While the resulting relevance judgments are still consistent,
    training with too many queries of this type may not be as useful for system
    performance
    since they are unlikely to capture the characteristics that the system needs to learn.

\textbf{Hallucination.} Sometimes the LLM inserts a detail into the query
    that is completely unrelated to anything in the source passage.
    For example, the query ``I seek news about Hong Kong's women's basketball team'' produced from a passage that includes information about a police basketball team and a mother who is a representative for a youth basketball team,
    but no mention of women's basketball teams,
    is a hallucination.
    This is a more problematic failure mode since the
    passage labeled relevant is not actually relevant to the query.

    \textbf{Overly broad.} Sometimes the LLM fails to detect that the non-relevant passage
    contains information that means that it is also relevant to the query. As with hallucination,
    this failure mode leads to inaccurate training data.
    Unlike hallucination,
    this type of failure is also found in the \msmarco training set,
    where negative examples were not necessarily judged by an assessor.

\subsection{Domain and Genre Shift}

A key claim of this paper is that building a CLIR training set
using the document collection of interest
will lead to better retrieval results
than just using a generic training collection such as the one underlying \msmarco.
CLIR evaluation collections over genres other than newswire are rare,
making this claim challenging to validate empirically.
To evaluate \jhpolo performance when the domain or genre does not match that of \msmarco,
we used the \hcthree collection~\cite{HC3}.
This collection comprises documents, queries, and relevance judgments
in Chinese and Persian.
The documents are Tweet reply threads of up to 100 Tweets in length.
Thus, the documents are short informal conversations --
very different from the web documents found in \msmarco.

\begin{figure*}[tb]
\begin{tabular}{rl}
 & - Taiwanese President Tsai Ing-wen's upcoming visit to Paraguay. \\
 & - The need for her to pass through the US to demonstrate her presence. \\ 
\textbf{Excluding addition} & - Making fun of Tsai Ing-wen's visit to Paraguay.  \\
 & - Suggestion to give Taiwan several billion in new Taiwan dollars to the country. \\
 & - Assertion that Taiwanese separatists will not be appeased without this. \\
\midrule
\multirow{2}{*}{\textbf{Including addition}}
 & - Taiwanese President Tsai Ing-wen's planned visit to Paraguay \\
 & - The possibility of Tsai offering monetary incentives to Paraguay during her visit \\

\end{tabular}
  \caption{Comparison of output when including the prompt addition
  ``No response should require the recipient to have seen the previous responses'' (below)
  or excluding it (above).}
  \label{fig:underspecified}
\end{figure*}

When shifting domains or genres, it may be necessary to re-engineer the prompt due to attributes of the data.
For instance,
we found that when using Tweets as our document collection, underspecificity was particularly egregious.
We experimented with prompt variants to ameliorate this problem.
We found that adding the sentence ``No response should require the recipient to have seen the previous responses''
was effective at eliminating many generic noun phrases,
which were at the core of most of the underspecificity problems.
Figure~\ref{fig:underspecified} shows output examples both with and without the additional sentence.
Its addition does not eliminate all underspecificity, but it greatly reduces it.
The figure illustrates three such queries where "the need for her," "to the country," and "without this" all lead to underspecified queries. This phenomenon is
not observed in the queries generated with the sentence.
Prompt generation is still a black art;
a variety of sentences conveying essentially the same requirement as this sentence
did not make an appreciable dent in underspecificity.

\section{Validation}

We performed two types of validation over the generated data. Section~\ref{sec:prompt_val} describes
the manual evaluation undertaken, while Section~\ref{sec:triple_val} describes an automated validation that
improves the quality of the training data.

\subsection{Prompt Validation}
\label{sec:prompt_val}

We manually annotated a small number of system outputs
to assess the quality of each prompt.
The assessor\footnote{Assessors were paper authors using Google passage translations. When there was questionable machine translation, a native speaker reviewed
the passage.} was provided with the two passages
and one of the the resulting queries.
Each such example was assigned to one of the following five categories,
based on whether the passage labeled relevant was truly relevant,
and whether the passage labeled non-relevant was truly not relevant:
\begin{itemize}
    \item Both assertions were correct.
    \item The assertion of relevance was incorrect.
    \item The assertion of non-relevance was incorrect.
    \item Both assertions were incorrect.
    \item The generated query was underspecified.
\end{itemize}
Treating each of these outcomes other than the first as erroneous,
the prompt shown in Figure~\ref{fig:prompt} applied to passage pairs selected as described in Section~\ref{sec:docselection} had an accuracy of 67\%
over 61 examples assessed.
While underspecified queries are probably not particularly useful training examples,
they also are unlikely to damage the training.
If the first and last outcomes are treated as correct,
accuracy rises to 72\%.

\subsection{Triple Validation}
\label{sec:triple_val}

After generation, we validate the triples with a multilingual cross-encoder reranking model. 
Validation is a important step to filter out triples that are likely to be hallucinations or that are overly broad.
One way to accomplish this validation is to use retrieval to ensure that
the positive passage is ranked first,
as was done in PROMPTAGATOR~\cite{promptagator}.
This was necessary in PROMPTAGATOR because negative passages were not included in its prompts.
\jhpolo prompts contain both a positive and a negative passage.
Therefore, our filtration process relies on the relative rankings of the two passages.
In particular, we only include a triple when the positive passage is \textit{more likely} to be relevant to the generated query than is the negative one;
this helps to ensure the integrity of the contrastive loss used during training.
Furthermore, we use a lower bound threshold on the difference between the two likelihoods
to ensure that the two are not too close in meaning with respect to the query. 

Specifically, let $F(q, p): \mathbb{R} \rightarrow \mathbb{R}$ be the cross-encoder model
that produces a real-valued score for a given query $q$ and passage $p$ pair.
For a given generated query,
positive and negative passage triple $(q, p_p, p_n)$, we consider the triple to be valid for training if 
\begin{equation}
    \frac{e^{F(q, p_p)}}{e^{F(q, p_p)} + e^{F(q, p_n)}} -
    \frac{e^{F(q, p_n)}}{e^{F(q, p_p)} + e^{F(q, p_n)}}
    > \tau
\end{equation}
where $\tau$ must be greater than 0;
otherwise, the negative passage is more likely to be relevant to the query than the positive passage.
We set $\tau$ to 0.15 in our experiments to eliminate noise from the training data. 

\section{Effectiveness Analysis}

In this section, we explore the effectiveness of \jhpolo-generated triples
by training retrieval models on them and comparing the performance
of those models over different evaluation datasets.
Our purpose here is not to try to match state-of-the-art retrieval effectiveness;
doing so is the purview of algorithms,
and thus outside of the scope of this paper.
We offer no new CLIR algorithms.
Rather, we show that \jhpolo-generated training data
are as good as machine-translated \msmarco data for collections that match those data well,
and superior to \msmarco data when the two diverge.

\subsection{Evaluation Collections}
\label{sec:generation}

We analyze the effectiveness of the \jhpolo methodology with two CLIR test suites -- 
TREC NeuCLIR 2022\footnote{The name of document collection is NeuCLIR~1. NeuCLIR 2022 refers to the evaluation suite that contains NeuCLIR~1 and the topics and relevance judgments developed for the TREC NeuCLIR Track 2022.}
and \hcthree~\cite{HC3}. 
Collection statistics appear in Table~\ref{tab:data-stats}.
These collections form the basis of our effectiveness analysis. 

The NeuCLIR 2022 dataset contains three sub-collections in Chinese, Persian, and Russian. 
Documents in NeuCLIR 1 are news articles extracted from Common Crawl News. 
The \hcthree dataset consists of Chinese and Persian Tweet reply threads each containing a root Tweet and up to 100 replies. 

When generating synthetic training data,
we draw passages from the target document collection;
therefore, passages are in the domain, genre, and language of the test collection.
Passage selection for the two collections differed
based on the quality of the written language in the passages.

For NeuCLIR 1, a positive passage was chosen randomly from all passages that exceeded a length requirement.
Length requirements, which were language-specific,
were set to the minimum document lengths imposed by the creators of the NeuCLIR 1 collection:
75 characters for Chinese, 100 characters for Persian, and 200 characters for Russian.
To identify a negative passage,
the positive passage was used as a query to search the collection,
and the resulting passages were ranked using BM25.
All BM25 scores were divided by the score of the positive passage.
The first passage of sufficient length
whose ratio of BM25 scores was less than 0.65
and where no other passages from that document
scored higher than 0.65 was selected as the negative passage. 

For \hcthree, the length minimums were reduced to 15 characters for Chinese and 25 for Persian.
However, a sample generation revealed that this process was insufficient for selecting Tweets
with enough content to generate understandable queries.
Consequently, we used 10,200 summaries from the WCEP multi-document summarization dataset~\cite{ghalandari2020}
as queries to select positive passages
(this dataset is  time-aligned with \hcthree,
and \hcthree topics tended to be event-inspired).
Since this summarization dataset is in English,
we use Google Translate to translate the summaries into the \hcthree languages to use as queries for BM25 retrieval.
For Chinese, \hcthree contains Tweets written in both Traditional and Simplified characters.
To retrieve Tweets in either character set,
translations were made into both Traditional and Simplified characters,
and the two translations were concatenated to form the queries.
Because these events were reported in the English media,
not all of them aligned well with topics in non-English Tweets.
To provide as much diversity as possible,
each relevant passage was uniquely paired with a single non-relevant passage;
thus no passage was paired with two different passages.
Another observed artifact was that re-Tweets greatly increased the presence of exact duplicate substrings.
This made it challenging for an LLM to create queries for which only one of the passages was relevant.
We handled this problem by imposing the longest common substring constraint.
While the BM25 ratio was still used,
we raised the threshold to 0.8 to create more passage pairs.
However, because BM25 gives great weight to unusual tokens,
URLs present in the Tweets introduced an unusual bias,
causing Tweets that were related by advertisements rather than by content to be chosen as pairs.
To handle this problem, we stripped URLs from all documents before passages were created.
In addition, two passages were paired only if they were both retrieved by the initial retrieval
and by the positive passage.
This led to the creation of fewer than 10,200 pairs.
Finally, the ``same document'' exclusion criterion used for the NeuCLIR 1 was dropped
since Twitter conversations are less coherent than Common Crawl News documents.

\subsection{Training Examples Generated}

Table~\ref{tab:generated-stats} summarizes the number of triples generated by GPT-3 Davinci-3. 
We generated roughly the same number of triples for all three sub-collections in NeuCLIR 1,
with Russian pairs having slightly more triples. 
Despite the prompt asking for five topics for each passage of the pair,
GPT-3 would not necessarily respond with the correct number,
and not all generated topics would pass the filter,
resulting in roughly eight queries per pair.
Generation for \hcthree is even more challenging,
with a fanout of around seven queries per passage pair. 

While enforcing unique passage pairs may seem desirable, 
this repetition of passage pairs is similar to the repetition of query and positive passage
that is found in \msmarco training triples. 
In fact, our generation process actually has less repetition than \msmarco.
In the small training triple file published by \msmarco,
there are roughly 100 negative passages associated with each query,
where the vast majority of the queries have only one positive passage.
Repetition of query-positive passage pairs is more than ten times that found in \jhpolo.
We argue that \jhpolo provides more diverse information in its triples
and thus has the potential to lead to better retrieval models. 

\subsection{Retrieval Models for Effectiveness Analysis}

We used two neural dense retrieval architectures as representatives  
for analyzing our methodology:
DPR-X~\cite{C3, mrtydi} and ColBERT-X~\cite{colbertx}. 
All models for each retrieval architecture are based on XLM-RoBERTa-base~\cite{conneau2019xlmr} 
started from the same checkpoint
and fine-tuned with a retrieval objective using English 
\msmarco for 200,000 update steps. 
We vary the source of the training data in the second stage fine-tuning, which consists of 1,245 steps.
This training scheme is designed to expose differences introduced
by a small amount of training data, rather than to train 
state-of-the-art systems.
Note that this training scheme does not include any advanced tricks such as iterative hard-negative mining~\cite{guu2020retrieval, xiong2020approximate}, in-batch negative sampling~\cite{dpr,repbert}, knowledge distillation~\cite{hofstaetter2020_crossarchitecture_kd}, etc. The training process here is a for demonstrating the relative effectiveness of \jhpolo as a training resource compared to \msmarco.

\jhpolo training triples were generated with the passage selection processes
for each evaluation collection outlined in Section~\ref{sec:generation}. 
GPT-3 Davinci-3 is prompted for queries using an English description along with a pair of
passages from the collection.
The generated queries, along with the passages, 
are passed through a cross-encoder trained on mMARCO~\cite{mmarco}
\footnote{\url{https://huggingface.co/cross-encoder/mmarco-mMiniLMv2-L12-H384-v1}}
for validation,
as described in Section~\ref{sec:triple_val}. 

To analyze the effectiveness of \jhpolo, we fine-tune the model in the second 
stage with the following regimens for comparison: 
\begin{itemize}

    \item \textbf{English}~(\textit{Eng.}). Continues fine-tuning the model with English \msmarco~v1.
    In this scenario, the model gains knowledge about non-English language during the initial training
    of the mPLM, but not during fine-tuning.

    \item \textbf{Translate}~(\textit{Trans.}). Fine-tuned with \msmarco~v1 documents that have been machine-translated into the language of the target document collection.\footnote{We used the NeuMARCO translation provided by the TREC NeuCLIR Track 2022. \url{https://ir-datasets.com/neumarco.html}}
    Queries remain in English, so the model is exposed to the CLIR task during continued fine-tuning,
    but the documents may contain translationese introduced by the machine translation system.
    This approach is also known as \textit{translate-train}~\cite{colbertx}. 

\end{itemize}

\begin{table}

\caption{Dataset statistics of NeuCLIR 2022 and \hcthree. 
}
\label{tab:data-stats}

\resizebox{\linewidth}{!}{
\centering
\begin{tabular}{c|rr|rr|rr}
\toprule
Collection & \multicolumn{2}{c|}{Chinese} & \multicolumn{2}{c|}{Persian} 
         & \multicolumn{2}{c}{Russian}   \\
Set   & \# Qry &  \# Docs 
	     & \# Qry &  \# Docs 
	     & \# Qry &  \# Docs 
	     \\
\midrule
NeuCLIR  
&     47 &  3,179,209     %
&   45 &  2,232,016     %
&   44 &  4,627,543     %
\\
\hcthree
&     50 &  5,584,146     %
&   50 &   7,335,221    %
&   - &   -    %
\\
\bottomrule
\end{tabular}
}

\end{table}

Following prior work in CLIR dense retrieval~\cite{colbertx, spladex}, 
we used the trained models to index the collections by separating the documents into overlapping
passages of 180 tokens with a stride of 90. 
Since both NeuCLIR and \hcthree consist of TREC-style topics, 
we concatenated the titles and descriptions as our search query;
these are the same queries used in the official NeuCLIR baseline runs for the reranking subtask. 
We evaluate the final retrieval effectiveness using nDCG@20
(the primary evaluation metric in TREC NeuCLIR)
and Recall at 100 (R@100).

\begin{table}[t]
    \caption{Statistics of the generation results. }
    \label{tab:generated-stats}
\centering
\begin{tabular}{l|rrr|r}
\toprule
         &  Passage & Generated &      Valid  &  Triples \\
         &    Pairs &   Triples &    Triples  &  Per Pair \\
\midrule
NeuCLIR  & & & & \\
\hspace{1em}Chinese  &   19,401 &   187,908 &   154,046  &  7.94 \\
\hspace{1em}Persian  &   19,432 &   180,174 &   153,933  &  7.92 \\
\hspace{1em}Russian  &   19,348 &   185,941 &   159,412  &  8.24 \\
\midrule
\hcthree & & & & \\
\hspace{1em}Chinese  &    9,766 &    86,532 &    68,679  & 7.03 \\
\hspace{1em}Persian  &   10,077 &    88,957 &    66,535  & 6.60 \\
\bottomrule
\end{tabular}

\end{table}
\begin{table*}[th]
\setlength\tabcolsep{0.4em}
\renewcommand{\b}[1]{\textbf{#1}}

\newcommand{\z}{$^*$} %
\renewcommand{\t}{$\dagger$} %

\caption{Retrieval Effectiveness. \z indicates significance with 95\% confidence against fine-tuning with English triples using paired t-tests with Bonferrini correction on three tests (over languages). \t indicates significance between \jhpolo and fine-tuning with translated triples using the same statistical test.  }
\label{tab:main-results}

\centering
\resizebox{\linewidth}{!}{
\begin{tabular}{p{4em}|lll|c|lll|c||ll|c|ll|c}
\toprule

& \multicolumn{8}{c||}{NeuCLIR 2022} & \multicolumn{6}{c}{\hcthree} \\
& \multicolumn{4}{c|}{nDCG@20} & \multicolumn{4}{c||}{R@100} & \multicolumn{3}{c|}{nDCG@20} & \multicolumn{3}{c}{R@100} \\
Triples
&  Chinese & Persian & Russian & Avg.
&  Chinese & Persian & Russian & Avg.
&  Chinese & Persian & Avg.
&  Chinese & Persian & Avg. \\
\midrule
\multicolumn{11}{l}{ColBERT-X} \\
\midrule
Eng.    &   0.155    &   0.131    &   0.227    &   0.171  &   0.236    &   0.290    &   0.290    &   0.272  &   0.198    &   0.196     &   0.197  &   0.361     &   0.368     &   0.364 \\
Trans.  &\b{0.216}\z &   0.220\z  &\b{0.267}\z &\b{0.234} &\b{0.320}\z &\b{0.389}\z &\b{0.325}\z &\b{0.345} &   0.208    &   0.254\z   &   0.231  &   0.385     &   0.400     &   0.393 \\
\jhpolo &   0.211    &\b{0.223}\z &   0.241    &   0.225  &   0.265    &   0.372    &   0.322    &   0.320  &\b{0.236}   &\b{0.270}\z  &\b{0.253} &\b{0.442}\z  &\b{0.419}    &\b{0.430}\\
\midrule
\multicolumn{11}{l}{DPR-X} \\
\midrule
Eng.    &   0.139    &   0.088    &   0.175    &   0.134  &   0.224    &   0.245    &   0.235    &   0.235  &   0.130    &   0.115     &   0.123  &   0.249     &   0.254     &   0.251 \\
Trans.  &   0.191\z  &\b{0.155}\z &\b{0.192}   &\b{0.179} &   0.280    &   0.317\z  &\b{0.278}   &   0.292  &   0.177    &   0.177     &   0.177  &   0.322     &   0.349     &   0.335 \\
\jhpolo &\b{0.192}\z &   0.132\z  &   0.181    &   0.168  &\b{0.294}\z &\b{0.343}\z &   0.277    &\b{0.305} &\b{0.240}\z &\b{0.269}\z\t&\b{0.255} &\b{0.500}\z\t&\b{0.483}\z\t&\b{0.491}\\

\bottomrule

\end{tabular}
}

\end{table*}

\subsection{Effectiveness on News Documents}

As presented in Table~\ref{tab:main-results},
both ColBERT-X and DPR-X benefit more from further fine-tuning with \jhpolo
than with more of the original English \msmarco for both nDCG@20 and Recall@100. 
Since \jhpolo is naturally cross-language,
a model trained on it could learn the definition of relevance directly from the target language pair. 
However, English \msmarco can only provide evidence on the relationship between queries and passages;
it cannot inform the CLIR system being trained about the target language.
This forces the model to rely solely on the multilinguality of the pretrained language model,
resulting in worse retrieval performance than if the training data encapsulated that information. 

By translating the \msmarco passages to the target language,
a model being trained can learn the cross-language relationships,
although the resulting passages will suffer from translationese. 
As repeatedly observed by prior work~\cite{colbertx, spladex, C3}, 
this translate-train approach provides state-of-the-art CLIR effectiveness when training only on \msmarco
but is dependent on the translation quality~\cite{colbertx}. 
When evaluating the models on NeuCLIR 2022,
whose documents are similar to \msmarco passages,
models trained with \jhpolo are only slightly worse than their translate-train counterparts.
These differences are not statistically significant,
indicating that the two approaches are similar
and neither consistently outperforms the other on all topics.
When evaluating on \hcthree,
which is a very different genre 
compared to \msmarco,
training with \jhpolo significantly outperforms translate;
we will discuss this outcome in detail in the next section.  

Comparing the two retrieval models,
DPR-X benefits from \jhpolo more than ColBERT-X does,
especially in the bottom part of the ranking (measured by recall). 
Since DPR-X summarizes each query and passage into a single vector,
it must rely on general semantics,
not on token-level matching.
Therefore, training with \jhpolo,
which contains queries that are only relevant to part of the positive passage
and do not necessarily have overlapping tokens,
improves DPR-X's ability to understand subtle differences between the passages.
In contrast, ColBERT-X focuses on more token-level cross-language alignments through translated passages,
directly enhancing its token-level matching.

However, triple quality is an artifact of the prompt used to generate it.
We value the diversity and the rich queries that our prompt can provide by generating topics instead of keywords.
This tendency implicitly benefits DPR-X more than ColBERT-X.
If one is only considering training a specific type of retrieval model,
the prompt can be adjusted to produce the kind of information the model most needs to
optimize its effectiveness. 

Again, we do not claim to reach the state-of-the-art CLIR effectiveness simply by training on \jhpolo;
such performance would require numerous optimizations,
such as using XLM-Roberta large rather than XLM-Roberta base,
fine-tuning for many steps beyond the reported two-stage regimen,
using in-batch negatives,
generating perhaps orders of magnitude more training examples,
and so on.
But what we do see here is that on a collection that is similar to the \msmarco collection,
\jhpolo generates training data that is on par with machine-translated \msmarco data.

\subsection{Effectiveness on Tweets}

When building a CLIR engine to search text that differs from the web articles that make up \msmarco,
training on \jhpolo provides dramatic improvements over \msmarco. 
When training with \jhpolo-generated triples on \hcthree,
both nDCG@20 and Recall@100 outperform translate-training with \msmarco. 
While translating the \msmarco passages into the target language
helps the retrieval model cross the language barrier,
the gap between the training genre and the \hcthree passages is still large.
\jhpolo fills this gap by directly exposing the model to Tweets during retrieval fine-tuning. 
Such exposure directly translates to effectiveness improvements across all regions of the ranked list. 

Interestingly, DPR-X is on par with, and sometimes better than, ColBERT-X
when trained with \jhpolo. 
This is unusual, as ColBERT-X generally outperforms DPR-X~\cite{C3}. 
We hypothesize that ColBERT-X requires more training data to learn how to match in a new genre 
than does DPR-X;
while ColBERT-X must adjust all term matches,
DPR-X only needs to adjust how its CLS token is created.
In this case, DPR-X is more efficient at absorbing the cross-language and cross-genre knowledge
provided by \jhpolo. 
Therefore, we argue that the smaller improvement in ColBERT-X when training on \jhpolo
is not necessarily the result of ineffective \jhpolo triples,
but of the nature of the retrieval model when searching across genre. 
Nevertheless, \jhpolo numerically improves ColBERT-X's performance on \hcthree,
although the difference is not statistically significant. 

Of particular note is the \jhpolo performance in Persian,
where in three of the four collection-retrieval system pairs \jhpolo outperforms translate,
one of which is a statistically significant difference.
Given that Persian is a lower resources language,
machine translation does not perform as well in general~\cite{colbertx}.
This indicates the using a high performing generative LLM may 
be able to provide better training data than machine translation.

\begin{figure*}

\newcommand{\multiline}[1]{\begin{minipage}[t]{\linewidth}\vspace{-0.7em}#1\vspace{0.3em}\end{minipage}}

\centering
\setlength{\arrayrulewidth}{0mm}

\resizebox{\linewidth}{!}{
\begin{tabular}{p{1.15\columnwidth}p{1.15\columnwidth}}
\toprule
\multiline{
\textbf{Passage A: }
\begin{CJK*}{UTF8}{gbsn}
1月25日,位于菲律宾阿尔拜省的马荣火山喷出火山灰。马荣火山位于菲律宾吕宋岛东南部的阿尔拜省，距菲首都马尼拉约330公里，海拔约2400米,是菲境内最活跃的火山之一。截至24日，已有超过7万人被疏散出马荣火山附近的危险区域。
\end{CJK*}  \\
\textit{Volcanic ash spews from Mayon Volcano in Albay, Philippines, on January 25. Mayon Volcano is located in Albay Province in the southeast of Luzon Island in the Philippines, locating about 330 kilometers away from Manila, the capital of the Philippines, and is about 2,400 meters above sea level. It is one of the most active volcanoes in the Philippines. As of the 24th, more than 70,000 people have been evacuated from the danger zone near Mayon Volcano.}
}
& 
\multiline{
\textbf{Passage B:}
\begin{CJK*}{UTF8}{gbsn}
当地时间13日早上5点43分，位于墨西哥市近郊的 \#波波卡特佩特火山 突然喷发，喷出大量火山灰，随即冒出白炽岩。 波波卡特佩特火山高达5426米，是墨西哥第二高峰,也是世界上最活跃的火山之一。
\end{CJK*} \\
\textit{At 5:43 a.m. local time on the 13th, the \#Popocatépetl volcano located in the outskirts of Mexico City suddenly erupted, spewing out a large amount of volcanic ash, followed by incandescent rocks. With a height of 5,426 meters, Popocatepetl is the second highest mountain in Mexico and one of the most active volcanoes in the world.}
}
\\
\midrule
\multiline{
\textbf{Passage A being relevant: } \\
\cmark~ The location of Mt. Mayon in the Philippines \\
\cmark~ The evacuation of more than 70,000 people from the area near Mt. Mayon \\
\cmark~ The elevation of Mt. Mayon \\
\cmark~ The activity of Mt. Mayon compared to other volcanoes in the Philippines \\
\cmark~ The date of the eruption of Mt. Mayon
}
& 
\multiline{
\textbf{Passage B being relevant: } \\
\cmark~ The location of Popocatépetl volcano near Mexico City \\
\cmark~ The elevation of Popocatépetl \\
\cmark~ The activity of Popocatépetl compared to other volcanoes in Mexico \\
\cmark~ The date of the eruption of Popocatépetl \\
\cmark~ The composition of the materials emitted by Popocatépetl during the eruption 
}
\\
\midrule \midrule
\multiline{
\textbf{Passage A: }
\begin{CJK*}{UTF8}{gbsn}
欢迎和我一起在美国大选中~\#成为选民。马上寻找你的投票站，看看你可以为谁投票。我推举我们敬爱的习近平总书记！希望你也能投他一票 %
\end{CJK*} \\
\textit{Join me in \#becoming a voter in the US election. Find your polling place now and see who you can vote for. I recommend our beloved General Secretary Xi Jinping! I hope you can vote for him too}
}
& 
\multiline{
\textbf{Passage B:} 
\begin{CJK*}{UTF8}{gbsn}
美国大选日定在礼拜二，是因为当时美国人多为新教徒农民，周日去过教堂，周一动身出发，周二到达投票站投票。由此，大选日历来是亲朋好友难得聚会的好日子，大家会带上面粉、白菜和肉馅，到投票站一起包饺子，包好了一边吃一边投。而饺子不咬开就不知道什么馅，也寓意...
\end{CJK*}  \\
\textit{The U.S. election day is set on Tuesday because Americans were mostly Protestant farmers at that time, they went to church on Sunday, leave by Monday, and arrived at polling stations by Tuesday to vote. Therefore, the general election calendar has always been a rare good day for relatives and friends to gather. Everyone will bring flour, cabbage and minced meat to the polling station to make dumplings together, and vote while eating. And you don't know the kind of stuffing of the dumplings until you take a bite, which also means...}
}
\\
\midrule
\multiline{
\textbf{Passage A being relevant: } \\
\cmark~ Endorsement of Xi Jinping \\
\cmark~ Inviting others to join in voting for a particular candidate \\
\xmark~ The importance of voting in the US election \\
\xmark~ The importance of collective voting and participation \\
\xmark~ The necessity of actively seeking out one's local voting station
}
& 
\multiline{
\textbf{Passage B being relevant: } \\
\cmark~ A detailed history of the US voting system \\
\cmark~ Chinese-American cultural customs \\
\cmark~ How US citizens of different religions view election day \\
\cmark~ Traditional Chinese foods associated with the US election \\
\cmark~ The importance of family and friends gathering on election day
}
\\
\bottomrule
\end{tabular}
}
    
\caption{Sample queries generated by \jhpolo. Text in italics is the translation of the corresponding Chinese Tweet. \cmark~and \xmark~indicate whether the generated query passed the cross-encoder filter. }
\label{fig:jhpolo-examples}
\end{figure*}

\subsection{Analysis of Examples}
Figure~\ref{fig:jhpolo-examples} presents two passage pairs
and the queries that GPT-3 Davinci-3 generated from them.
Each passage pair is connected by an identifiable thread
(\textit{eruptions} in the top of the figure;
\textit{voting} in the bottom).
Because of the way they were selected,
these passages tend to contain more information than a randomly selected Tweet. Many of the topics for the eruptions are similar,
but are specific to the eruption mentioned in the positive passage.
We do see that occasionally there is a query for which there is no further information in the passage,
such as the location of Popocatépetl. 

Because the bottom passages are less formal,
the queries are more general.
In particular, 
the bottom Passage A produces only two queries;
the other three were filtered out during the validation step
and are marked with an \xmark.
The remaining queries are supported by the passage.
In the queries for Passage B,
the first one clearly identifies a topic in the passage;
however, the third query concerning religions is not well supported,
as the passage does not explain the viewpoints of Protestant farmers.
While one might question the connection between traditional Chinese food and US elections,
the passage does include information on that,
and the LLM captures it well.

\section{Cost}

GPT-3 is not free;
the cost of producing a CLIR training collection using \jhpolo
depends on the size of the collection and the cost of GPT-3 per request.
At this writing, GPT-3 Davinci-3 (the most capable model)
costs us US\$0.02 per 1000 subword tokens
(the sum of the number of tokens in the prompt and in the output).
Subwords are produced by the GPT-2 tokenizer,\footnote{\url{https://beta.openai.com/tokenizer}}
which is similar to SentencePiece.\footnote{\url{https://github.com/google/sentencepiece}}
Thus, our training corpus built on the Chinese NeuCLIR collection cost us about US\$400 to produce,
while Persian and Russian cost about 20\% more.
GPT-3 throughput has been about two prompts per second with ten concurrent processes,
allowing us to create the collections in about 16 hours.

The cost per 1000 training examples for the NeuCLIR collections averaged US\$3 for the prompt shown in Figure~\ref{fig:prompt}.
The number of requests is the same as the number of passage pairs shown in Table~\ref{tab:generated-stats}. 
While the cost does add up, it is orders of magnitudes cheaper than producing a dataset annotated by humans, 
such as \msmarco.

\section{Conclusions}

This paper introduces the \jhpolo CLIR training set creation methodology,
which selects a positive and negative passage from the target document collection
and uses a generative large language model to synthetically generate one or more queries per passage pair.
The methodology suggests that random selection of positive passages works well for high quality texts,
and shows how to select passages with meaningful content
for noisier texts.
It allows negative examples to be selected before the query is generated,
thus providing some control over the quality and difficulty of the training collection.
It demonstrates effective prompts that describe the desired output
without the need for exemplars.
We find that GPT-3 Davinci-3 can generate sufficiently good
queries so that the resulting training triples can train a CLIR retrieval model as effectively as \msmarco.
In addition, the further the genre of the document collection is from Bing web passages,
the more effective the synthetically generated data is.
Thus, \jhpolo offers a pathway to automatically creating an effective CLIR training set
for any text corpus of interest.

\bibliographystyle{ACM-Reference-Format}
\bibliography{ms}

\end{document}